\documentclass{emulateapj}
\usepackage{threeparttable}
\usepackage[backref,breaklinks,colorlinks,citecolor=blue]{hyperref}

\newcommand{\source}{PSR$\;$J1846$-$0258}

\shorttitle{On the braking index of \source{}}
\shortauthors{Archibald et al.}

\begin{document}

\title{On the Braking Index of the Unusual High-$B$ Rotation-Powered Pulsar \source{}}

\author{R. F. Archibald\footnotemark[1], V. M. Kaspi\footnotemark[1], A. P. Beardmore\footnotemark[2], N. Gehrels\footnotemark[3], \& J. A. Kennea\footnotemark[4]}

 \footnotetext[1]{Department of Physics, McGill University, Montreal QC, H3A 2T8, Canada}
 \footnotetext[2]{Department of Physics and Astronomy, University of Leicester, University Road, Leicester LE1 7RH, UK}
 \footnotetext[3]{Astrophysics Science Division, NASA Goddard Space Flight Center, Greenbelt, MD 20771 USA}
 \footnotetext[4]{Department of Astronomy and Astrophysics, 525 Davey Lab, Pennsylvania State University, University Park, PA 16802, USA}

\begin{abstract}
\source{} is an object which straddles the boundary between magnetars and rotation powered pulsars. Though behaving for many years as a rotation-powered pulsar, in 2006, it exhibited distinctly magnetar-like behavior -- emitting several short hard X-ray bursts, and a flux increase. Here we report on 7 years of post-outburst timing observations of \source{} using the {\it Rossi X-ray Timing Explorer} and the {\it Swift} X-ray Telescope. We measure the braking index over the post-magnetar outburst period to be  $n=2.19\pm0.03$. This represents a change of $\Delta n=-0.46\pm0.03$ or a 14.5$\;\sigma$ difference from the pre-outburst braking index of $n=2.65\pm0.01$, which itself was measured over a span of 6.5 yr. So large and long-lived a change to a pulsar braking index is unprecedented and poses a significant challenge to models of pulsar spin-down. 
\end{abstract}

\section{Introduction}
Many of the quoted properties of pulsars, such as the surface magnetic field, the characteristic age, and the spin-down luminosity, are based on the assumption that pulsars are well modeled as a magnetic dipole in a vacuum.
One of the ways we have to test the validity of this assumption, and by doing so probe the emission mechanisms of pulsars,  is by measuring the change of a pulsar's spin-down rate over time.
This is expected to behave following a power law,
\begin{equation}
\dot{\nu}=-K\nu^{n},
\end{equation}
where $\nu$ is the spin frequency of the pulsar and $n$ is referred to as the `braking index'.
In the canonical case of a rotating magnetic dipole in a vacuum, the braking index is expected to be 3 \citep[e.g.][]{1977puls.book.....M}.

Observationally, the braking index is measured by means of observing a gradual change in $\dot{\nu}$, the frequency derivative, and expressing
\begin{equation}
n=\frac{\ddot{\nu}\nu}{\dot{\nu}^2}.
\end{equation}
Measuring braking indices for pulsars has proven to be difficult, as young pulsars typically exhibit large amounts of timing noise which can contaminate measurements of $\ddot{\nu}$ \citep[e.g.][]{2010MNRAS.402.1027H}.
As yet, only eight pulsars have measured braking indices \citep[see][and references therein]{2012MNRAS.424.2213R, 2011ApJ...741L..13E} which range from 0.9 \citep{2011ApJ...741L..13E} to 2.91 \citep{2011MNRAS.411.1917W}.
 A braking index of $-1.5$ was reported for PSR$\;$J0537$-$6910  but the timing behavior is complex, and dominated by glitches as discussed in \cite{2006ApJ...652.1531M}.

\source{} is a $\sim$800 year old pulsar located in the Kesteven 75 supernova remnant \citep{2000ApJ...542L..37G}.
It has a rotation period of $\sim$327 ms and is one of the youngest known pulsars.
For the majority of its observed lifetime, \source{} behaved as if it were a typical rotation-powered pulsar, with its X-ray emission being much less than the luminosity explainable by its spin-down power.
Curiously,  however it has no detectable radio emission \citep{2008ApJ...688..550A}.
\source{} is also one of the eight pulsars with a measured braking index, observed to be 2.65$\pm$0.01 from 2000 to 2006 \citep{2006ApJ...647.1286L}.

In 2006, \source{} underwent a rare event - its  pulsed X-ray flux increased dramatically, it had a large glitch, and emitted several magnetar-like bursts \citep{2008Sci...319.1802G, 2008apj...678.1L43K, 2009A&A...501.1031K}.
\source{} remains the only seemingly rotation-powered pulsar to display such distinctly magnetar-like behavior,  making it an interesting transition object between the two classes.

After this magnetar-like outburst, \source{} went back to manifesting itself as a rotation-powered pulsar \citep{2011ApJ...730...66L}.
However, after timing the source for more than two years post-outburst, \cite{2011ApJ...730...66L} measured a braking index of $n=2.16\pm0.13$ during this period, a value inconsistent with the braking index measured prior to the outburst.
While the braking index is expected to change on a timescale of thousands of years \citep[see e.g.][]{2006ApJ...643.1139C,2015MNRAS.446.1121G}, such a sudden change is unexpected in the standard models.

Here we report a further five years of X-ray timing observations of \source{}, for a total of  seven years after the magnetar-like outburst.
We show that the braking index is consistent with the post-outburst measurement of \cite{2011ApJ...730...66L}, and inconsistent with that prior to the outburst.
This indicates that the  2006 magnetar-like outburst resulted in a persistent change in the braking index in the source.

\section {Observations and Analysis}

\subsection{RXTE}
In this work, we analyze observations of \source{} from the Proportional Counting Array (PCA) aboard the {\it Rossi X-ray Timing Explorer} ({\it RXTE}) from  January 2008 until the decommissioning of {\it RXTE} in 2011 December.
The PCA consists of five collimated xenon/methane multianode proportional Counter units (PCUs) which are sensitive to photons in the 2--60$\;$keV range \citep{1996SPIE.2808...59J, 2006ApJS..163..401J}.
The PCA was operated in ``Good Xenon'' mode, which provides 1-$\mu$s resolution for photon arrival times.

Observations were obtained from the HEASARC archive and barycentered to the \
location  of \source{}, $RA= 18^h46^m24.94^s$, $DEC=-02^\circ  58' 30.1''$ \citep{2003ApJ...582..783H} using the $barycorr$ tool in HEASOFT $v6.16$.
Observations were filtered to remove non-astrophysical events using $xtefilt$.
In order to maximize the signal-to-noise ratios of pulse profiles so as to minimize uncertainties on resulting pulse times-of-arrival (TOAs) (see \S\ref{sec:timing}),we used events from all layers of the then-operational PCUs.

In total  363 {\it RXTE} observations providing $\sim0.9$ Ms of exposure time were analysed in this work spanning January 2008 to December 2011.  Observations taken within 2 days of each other were merged, resulting in 177 TOAs for a typical exposure time of 5-ks per TOA.

\subsection{Swift XRT}
We began observing \source{} with the {\it Swift} X-ray Telescope (XRT) on  2011 July 25 as part of a campaign to monitor several magnetars \citep[see e.g.][]{ 2013Natur.497..591A, 2014ApJ...783...99S, 2015ApJ...800...33A}.
The {\it Swift} XRT is a Wolter-I telescope with a e2v CCD22 detector, sensitive in the $0.3-10\;$keV range.
The XRT was operated in Windowed-Timing (WT) mode for all observations.
This gave a time resolution of $1.76\;$ ms.

Level 1 data products were obtained from the HEASARC \emph{Swift} archive, reduced using the $xrtpipeline$ standard reduction script, and barycentered to the location  of \source{}, using HEASOFT $v6.16$.
Individual exposure maps, spectra, and ancillary response files were created for each orbit and then summed.
We selected only Grade 0 events for spectral fitting as higher Grade events are more likely to be caused by background events \citep{2005SSRv..120..165B}.
To maximize the signal-to-noise ratios of pulse profiles so as to minimize uncertainties on resulting pulse times-of-arrival (see \S\ref{sec:timing}), only photons from 2.7--10$\;$keV were used.

To investigate the flux and spectral evolution of \source{}, a circular region having a 10-pixel radius  centered on the source was extracted.
As well, an annulus of inner radius 75 pixels and outer radius 125 pixels centered on the source was used to extract background events.

In total 66 XRT observations totaling 541 ks of exposure time were analyzed in this work.
Observations taken less than 5 days apart were grouped to extract a single TOA yielding 47 TOAs, with a typical exposure time of 10 ks per TOA.

\section{Timing Analysis}
\label{sec:timing}
\subsection{Phase-Coherent Timing Analysis}

TOAs for all {\it RXTE} and {\it Swift} observations were extracted using a Maximum Likelihood (ML) method as described in  \cite{2009LivingstoneTiming} and \cite{2012ApJ...761...66S}.
The ML method compares a continuous model of the pulse profile to the photon arrival times obtained from a single observation.
In order to create the continuous model of the pulse profile, first we create a high signal-to-noise template profile by folding many observations together using a whitened timing solution.
For {\it RXTE}, the template was derived from folding  all pre-outburst observations, and for {\it Swift}, using all the observations.
Separate templates were used for the {\it RXTE} and {\it Swift} observations to account for differences in the responses of the telescopes.
In both cases, a continuous model of the profile was created by fitting the high signal-to-noise template with a Fourier model using the first two harmonics.
Two harmonics were chosen to optimally describe the pulse shape, as determined by the h-test \citep{1989A&A...221..180D, 2015ApJ...807...62A}.

These TOAs were fitted to a timing model in which the phase as a function of time $t$ can be described by a Taylor expansion:
\begin{equation}
\phi(t) = \phi_0+\nu_0(t-t_0)+\frac{1}{2}\dot{\nu_0}(t-t_0)^2+\frac{1}{6}\ddot{\nu_0}(t-t_0)^3+\cdots
\end{equation}
where $\nu_0$ is the rotational frequency of the pulsar at time $t_0$.
This was done using the TEMPO2 \citep{2006MNRAS.369..655H} pulsar timing software package.

\begin{figure}
\centering
\includegraphics[width=\columnwidth]{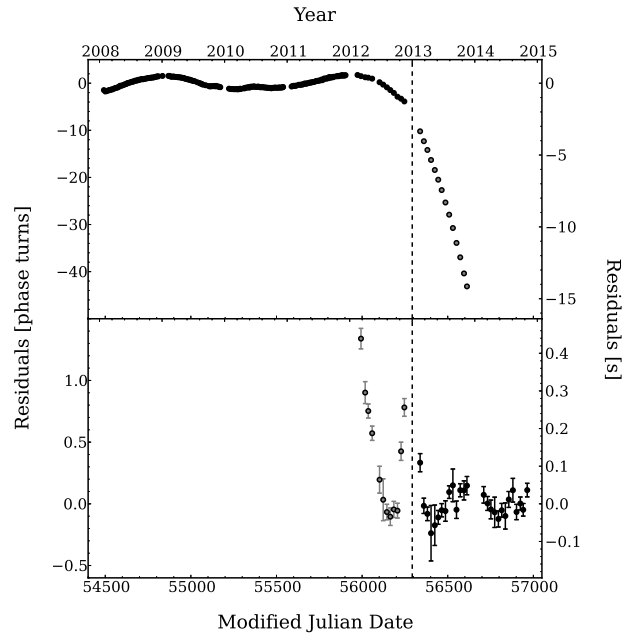}
\caption[Residuals]{Timing residuals of \source{} from MJD 54492-56880 (post-outburst) for the solutions presented in Table~\ref{tab:timing}. The top panel shows the residuals of Solution 1. The bottom panel shows the residuals of Solution 2. The vertical dashed line indicates where there is a phase ambiguity; see \S\ref{sec:timing} for details. Note that Solution 1 is fitted only to data before the phase ambiguity, and Solution 2 only to those after.}
\label{fig:resplot}
\end{figure}

In Figure~\ref{fig:resplot} we show the timing residuals in the range MJD 54492 to 56880, the period after the magnetar-like outburst and glitch recovery have relaxed.
For details about the glitch and the glitch recovery, see  \cite{2009A&A...501.1031K}  and \cite{2010ApJ...710.1710L}.

Finding a single phase-coherent solution over the entire seven-year post-outburst data set is not possible due to a phase ambiguity during the Sun constraint period from MJD 56246 to 56338.
This is indicated in Figure~\ref{fig:resplot} by a dashed vertical line.
We were able to find two phase-coherent solutions, one before this Sun constraint and one after.
The two timing solutions are presented in Table~\ref{tab:timing}.

This loss of phase coherence could be due either to a glitch, or to timing noise.
Fitting for a glitch during the Sun-constraint period using a timing solution up to $\ddot{\nu}$ yields $\Delta\nu/\nu = 5.7\pm0.5\times 10^{-8}$ and $\Delta\dot{\nu}/\dot{\nu}= -2.5\pm0.4\times 10^{-4}$  over the {\it Swift} campaign.
Fitting using both the {\it RXTE} and {\it Swift} data sets gives glitch parameters ranging from $\Delta\nu/\nu$ of $-9\times 10^{-8}$ to $1.7\times 10^{-7}$.
We note that these values vary based on the time-span fit and the number of frequency derivatives used in the fit.
Finally, we note that fitting a continuous solution over the gap yields comparable residuals to the glitch fits.
Thus we do not need to invoke a sudden glitch to explain the timing behavior of \source{} at this epoch.

\begin{table}
\begin{center}
\caption{Phase-Coherent Timing parameters for \source.}
\label{tab:timing}
\begin{tabular}{ll}

\hline
\multicolumn{2}{c}{First Phase-coherent Solution} \\
\hline
Dates (MJD)         & 54492.0-56246.7 \\
Dates               & 2008 Jan 27 - 2012 Nov 15 \\
Epoch (MJD)         & 55369.00000\\
$\nu\;$ (s$^{-1}$)            & 3.059 040 903(4)\\
$\dot{\nu}\;$ (s$^{-2}$)            & $-6.651 31(1)\times 10^{-11}$\\
$\ddot{\nu}\;$ (s$^{-3}$)            & $2.937(8)\times 10^{-21}$ \\
rms residual (ms) & 304.4\\
rms residual (phase) & 0.931\\
Braking index, $n$                              & 2.031(6)\\
\hline
\multicolumn{2}{c}{Second Phase-coherent Solution } \\
\hline
Dates (MJD)        & 56338.7-56964.20\\
Dates               & 2013 Feb 15 - 2014 Nov 03 \\
Epoch (MJD)         & 56651.00000\\
$\nu\;$ (s$^{-1}$)            & 3.051 693 972(3)\\
$\dot{\nu}\;$ (s$^{-2}$)            & $-6.613 49(2)\times 10^{-11}$\\
$\ddot{\nu}\;$ (s$^{-3}$)            & $3.30(4)\times 10^{-21}$\\
rms residual (ms) & 32.4\\
rms residual (phase) & 0.099\\
Braking index, $n$             & 2.30(3)\\
\hline
\end{tabular}

Note: Figures in parentheses are  the nominal 1$\sigma$ \textsc{tempo2} uncertainties in the least-significant digits quoted.
\end{center}
\end{table}

\subsection{Partial Phase-Coherent Timing Analysis}
Measurements of $\ddot{\nu}$ can be susceptible to contamination from timing noise \citep[e.g.][]{2010MNRAS.402.1027H}.
To mitigate this effect, we fit small segments of data to make local measurements.
For all methods presented below, relative pulse numbers were fixed to those given by the fully phase-coherent timing solution.
No timing solution was fit overlapping the phase ambiguity.

For each small segment of data, using the established pulse numbers, TOAs were fit to a timing solution consisting of only $\nu$ and $\dot{\nu}$.
The time spans were determined by allowing a maximum $\chi^2_\nu$ of $\sim$1 and the condition that there was no apparent-by-eye red-noise signal in the residuals.
When this condition was met, we moved over by half the number of TOAs in that solution, and fit again until the criteria were met.
We did not allow a solution to span over a Sun-constraint period.

\begin{figure}
\centering
\includegraphics[width=\columnwidth]{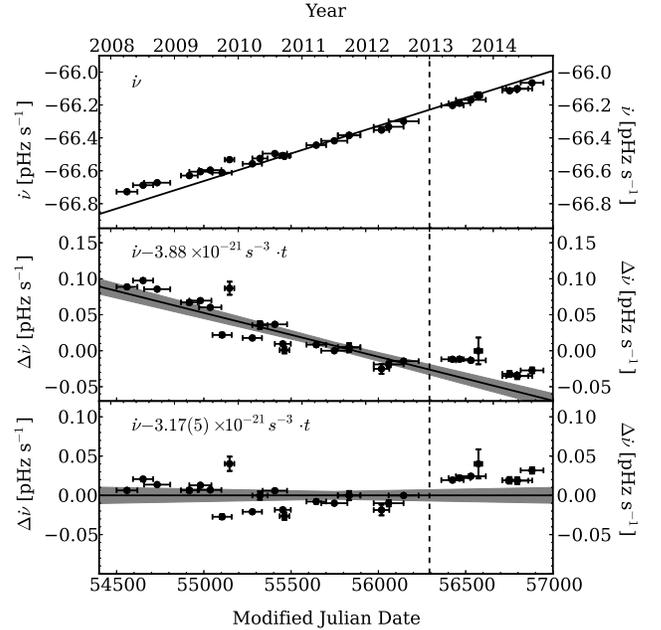}
\caption[nudot]{$\dot{\nu}$ measurements for \source{} from MJD 54492-56880. The top panel shows  the measured $\dot{\nu}$. The solid black line shows the pre-outburst $\ddot{\nu}$ of $3.88\times 10^{-21}\;$s$^{-3}$.The middle panel shows the same data subtracting the pre-outburst $\ddot{\nu}$.
The black line in this panel shows the difference between the pre-outburst measurement, and the best-fit post-outburst $\ddot{\nu}$ of $3.17\pm0.05\times 10^{-21}\;$s$^{-3}$. The gray shaded region shows the 1-$\sigma$ bounds on this determined from a bootstrap analysis to the full data set as described in the text.
The bottom panel shows the $\dot{\nu}$ residuals after subtracting the best-fit slope from above.
The vertical dashed line indicates where there is a phase ambiguity; see \S\ref{sec:timing} for details.}
\label{fig:nudot}
\end{figure}

In Figure~\ref{fig:nudot}, we show these measurements of $\dot{\nu}$ over the data set. The top panel shows  $\dot{\nu}$  over time. The middle panel shows  $\dot{\nu}$ over time subtracting a constant slope consisting of the pre-outburst braking index, 2.65$\pm$0.01 \citep{2011ApJ...730...66L}.
Note the clear linear trend in the middle panel indicating that the pre-outburst braking index does not describe the data well.

We fit a slope to the post-outburst $\dot{\nu}$ in order to obtain a measurement of $\ddot{\nu}$, and thus a braking index.
The timing measurements of \source{} have a scatter larger than would be suggested by their formal errors, therefore we use a bootstrap method.
The bootstrap method is robust for error estimation  when only a small number of measurements are available \citep{efron1979} and the formal uncertainties are thought to not fully describe the data.

For the full post-outburst data set, this yielded a measurement of $\ddot{\nu}=3.17\pm0.05\times10^{-21}\;$s$^{-3}$ corresponding to a braking index of $n=2.19\pm0.03$ for the bootstrap method.
The residuals of this fit can be seen in the bottom panel of Figure~\ref{fig:nudot}.

In order to verify that the phase ambiguity between the two timing solutions presented in Table~\ref{tab:timing} does not affect our result, we split the data into the corresponding two segments.
Fitting from MJD 54492.0-56246.7 gives a braking index of $n=2.09\pm0.05$.
Fitting MJD 56338.7-56880.5 gives $n=2.23\pm0.07$.
These two segments gave consistent slopes at the 1.4-$\sigma$ level, and are both inconsistent with the pre-outburst braking index.
This gives us confidence that the measured post-outburst braking index of $n=2.19\pm0.03$ represents a long-lived change in the braking index of $\Delta n=-0.46\pm0.03$, a 14.5$\;\sigma$ difference.

\subsection{Timing Noise}
In order to quantify the effect of timing noise that could be contaminating the measurement of the braking index, we fit a timing solution consisting of a frequency and three frequency derivatives for each year, ending a solution at times of a glitch, or the start of Sun-constraint.

Following the method of \cite{2011ApJ...730...66L}, we measure the quantity
\begin{equation}
\label{eqn:deltanudddot}
\Delta_{\dddot{\nu}}\equiv \log\left(\frac{1}{24}\frac{\left|\dddot{\nu}\right|t^4}{\nu}\right)
\end{equation}
where $t$ is the length of time over which the solution was fit, $\sim2.5\times10^{7}\;$s.
This is analogous to the $\Delta_8$ parameter of \cite{1994ApJ...422..671A} where $\Delta_8$ is used as an estimation of the contributions of $\ddot{\nu}$ to the accumulated phase deviation of the pulsar.
As $\ddot{\nu}$ is physically relevant in timing measurements of \source{}, we use $\Delta_{\dddot{\nu}}$ as an estimate of the phase contamination from $\dddot{\nu}$ and higher order effects.

\begin{figure}
\centering
\includegraphics[width=\columnwidth]{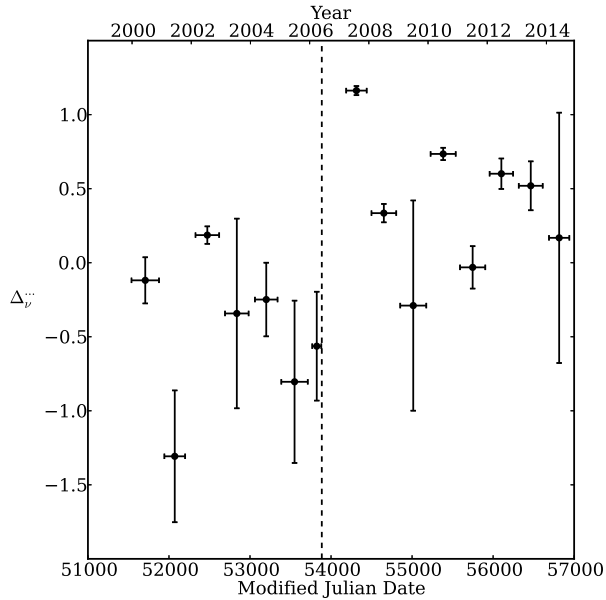}
\caption[timing_noise]{Timing noise in \source{} over 15 years of X-ray timing as described by the $\Delta_{\dddot{\nu}}$ parameter, see equation~\ref{eqn:deltanudddot} . The vertical dashed line indicates the epoch of the outburst.}
\label{fig:timingnoise}
\end{figure}

In Figure~\ref{fig:timingnoise}, we show $\Delta_{\dddot{\nu}}$ over the 15 years of timing of this source.
While the scatter is high, $\Delta{\dddot{\nu}}$ shows a possible increase for the period following the  magnetar-like outburst in 2006.

Before the outburst, eg. from 2000 to 2006, the weighted mean was $\Delta_{\dddot{\nu}}=0.1\pm0.2$.
For the first period after the outburst, 2007, $\Delta_{\dddot{\nu}}=1.16\pm0.03$, substantially higher than at any other time.
After this, the timing noise decreased to a level that is marginally higher than the pre-outburst noise, with the weighed mean of $\Delta_{\dddot{\nu}}=0.6\pm0.2$ from 2008-2014.
Thus, the level of timing noise clearly increased following the magnetar-like outburst but appears to be relaxing back to the pre-outburst level on a time scale of several years.

\section{Radiative properties}
\subsection{Spectral Analysis}

{\it Swift} XRT spectra were extracted from the selected regions using {\tt extractor}, and fit using {\tt XSPEC} package version 12.8.2\footnote{http://xspec.gfsc.nasa.gov}.
Spectral channels were grouped to 1 count per bin, and fitted using $cstat$ minimization.
The spectrum was fit with a photoelectrically absorbed power law.
Photoelectric absorption was modeled using {\tt XSPEC} {\tt tbabs} with abundances from \cite{2000ApJ...542..914W}, and photoelectric cross-sections from  \cite{1996ApJ...465..487V}.
Due to both the nature of the windowed timing read-out mode of the XRT, and the fact that the XRT point spread function is comparable to the size of the bright, central region of the nebula, we are unable to separate the flux coming from the pulsar itself from the bulk of the pulsar wind nebula which surrounds it.

As all of the {\it Swift} observations had consistent flux and spectral parameters, we co-fit all observations simultaneously.
This yielded a best-fit model with $N_H=(4.43\pm0.05)\times10^{22}$ cm$^{-2}$ and $\Gamma=1.80\pm0.02$.
We note that the best-fit power-law index is consistent with that of the pulsar wind nebula reported by \cite{2008apj...678.1L43K} and \cite{2008ApJ...686..508N},  as well as that reported in the 20--300$\;$keV range using {\it INTEGRAL} \citep{2009A&A...501.1031K}.

The absorbed 0.5--10$\;$keV X-ray flux measured over the {\it Swift} campaign of the combined pulsar and pulsar-wind-nebula was $(2.04\pm0.02)\times10^{-11}$ erg cm$^{-2}$ s$^{-1}$.
In {\it Chandra} observations taken 2000 \citep{2008ApJ...686..508N} and 2009 \citep{2011ApJ...730...66L}, the absorbed 0.5--10 keV flux from the combined pulsar and pulsar-wind-nebula were (1.81$\pm$0.03)$\times10^{-11}$ erg cm$^{-2}$ s$^{-1}$ and (1.73$\pm$0.07)$\times10^{-11}$ erg cm$^{-2}$ s$^{-1}$ respectively.
While formally, our measured {\it Swift} flux and the archival {\it Chandra} fluxes are inconsistent, the cross-calibration between X-ray instruments is only accurate to the $\sim$10\% level; see \cite{2011A&A...525A..25T}.
Therefore we find no evidence of a changing flux for the system to the level of instrumental uncertainties.

\subsection{Pulse Profile Analysis}
\label{subsect:profile}

To look for changes in the pulse profile, we folded each observation using 16 phase bins.
Each profile was compared to the high signal-to-noise-ratio pulse template described in \S\ref{sec:timing} by subtracting a fitted DC offset, and using a multiplicative scaling factor to minimize the difference between the template and scaled profile as determined by a $\chi^2$ minimization.
For both {\it RXTE} and {\it Swift}, all profiles are consistent with the respective telescope's standard template.

\begin{figure}
\centering
\includegraphics[width=\columnwidth]{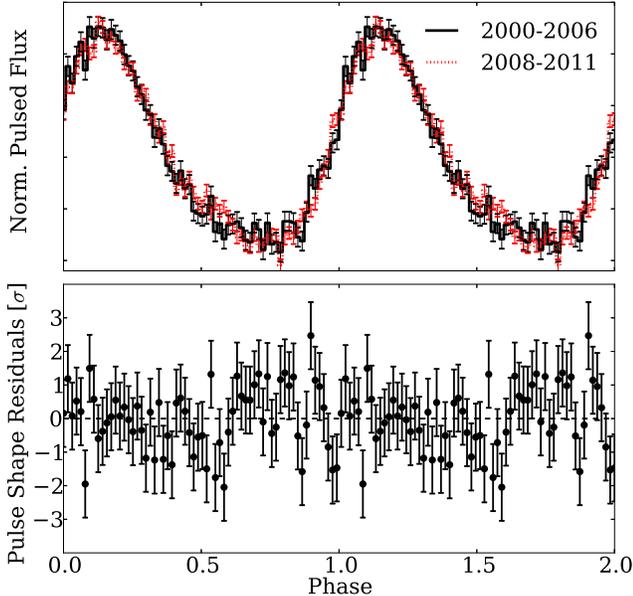}
\caption[flux]{Normalized {\it RXTE} pulse profiles of \source{}. The solid black profile shows the profile from January 2000 to March 2006, just before the 2006 outburst. The red dotted profile shows the profile from 2008 January to 2011 December. The bottom panel shows the residual difference between the two profiles. The residuals have $\chi^2_\nu/(dof)=0.988/(62)$ indicating the profiles are statistically identical.}
\label{fig:prof}
\end{figure}

It has been shown previously that  \source{} exhibited no significant change in its X-ray pulse profile during the magnetar event \citep{2009A&A...501.1031K, 2010ApJ...710.1710L}.
For {\it RXTE} we now have comparable data from both before and after the magnetar-like outburst.
This allowed us to search for long-term  lower-level changes in the pulse profile.
To do so, we combined all observations for which we had a valid timing solution into a two high signal-to-noise-ratio profiles with 64 phase bins representing the pulse profile before and after the magnetar-like outburst.
To do this, we fitted each year of TOAs to a timing solution, using as many frequency derivatives as necessary to whiten the residuals.
Each year of data was then folded into a 64-bin profile, and aligned  with other years' profiles using cross-correlation.
This resulted in two high signal-to-noise-ratio profiles: the first using 918$\;$ks of exposure time from January 2000 to March 2006, and the second using 871$\;$ks from January 2008 to December 2011.
These two normalized, DC-subtracted, high signal-to-noise profiles are shown in Figure~\ref{fig:prof}, as well as the difference between them.
The residuals have $\chi^2_\nu/(dof)=0.988/(62)$ indicating the profiles are statistically identical.
This is consistent with the lack of profile change reported by \cite{2009A&A...501.1031K} and \cite{2011ApJ...730...66L}.

\subsection{Burst Search}
All {\it Swift} observations were searched for magnetar-like bursts by binning the source region light curves into 0.01-s, 0.1-s, and 1.0-s bins.
The counts in each bin were compared to the mean count rate of its Good Timing Interval (GTI), assuming Poisson statistics, similar to the methods described by \cite{2011ApJ...739...94S}.
We found no significant bursts in the {\it Swift} observations.

\begin{figure}
\centering
\includegraphics[width=\columnwidth]{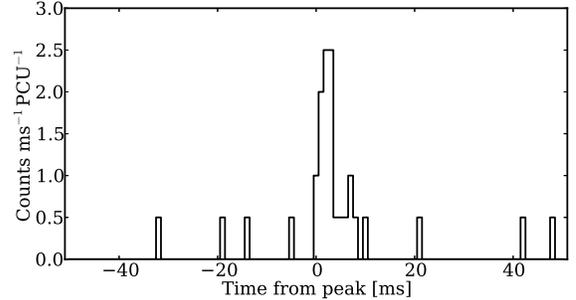}
\caption[burst]{Burst from the direction of \source{} on MJD 55070. The time series is binned with 1-ms time resolution and covers the full 2--60$\;$keV range of the PCA of both operational PCUs.}
\label{fig:burst}
\end{figure}

For the {\it RXTE} PCA, due to the background being highly variable, each 60-second interval was treated similarly to a {\it Swift} GTI.
An additional constraint was placed on the PCA data that a putative burst must be detected in all operational PCUs to be considered real.
We find a  previously unreported burst on MJD 55070, 27 August, 2009.
This burst has a $T_{90}$, the time duration in which 90$\%$ of a burst's fluence is collected, of $7\pm1\;$ms and a fluence of $12\pm3$ counts per PCU ($24\pm5$ total counts).
This corresponds to a false alarm probability of $\sim10^{-20}$ for the observation.
The burst is shown in Figure~\ref{fig:burst}.

We note, however, that the field of view contains other known magnetars including AX~J1845.0$-$0300 \citep{1998ApJ...503..843T} located $0.38\,^{\circ}$ from the center of the pointing, and 1E~1841$-$045 \citep{1997ApJ...486L.129V}, located $2.3\,^{\circ}$  from the center of the pointing.
As 1E 1841$-$045 is an active and frequent burster \citep[e.g.][]{2011ApJ...740L..16L, 2015arXiv150503570A}, it is possible that the burst originated from this source.
While we cannot exclude the possibility that this burst originated from \source{}, we note that there is no change to either the radiative properties, or timing behavior at these epochs to within our measurement uncertainties.

\section{Discussion}
\label{sec:discussion}
We have presented seven years of post-outburst timing of \source{} in which we measure the braking index to be $n=2.19\pm0.03$.
This is discrepant at the $14.5\sigma$  level from the pre-outburst braking index of $n=2.65\pm0.01$  \citep{2006ApJ...647.1286L}.
We note that this measurement is made over a comparable span of time to that over which the pre-outburst braking index was measured.

Only one other rotation-powered pulsar has had a radiative change associated with a glitch: PSR$\;$J1119$-$6127.
Following a glitch in 2007, the radio pulse profile changed from single- to double-peaked.
This double-peaked profile was only seen once, during the first post-glitch observation of the pulsar, and had returned to the single-peaked profile by the next observation \citep{2011MNRAS.411.1917W}.
It appears that PSR$\;$J1119$-$6127 may have undergone a change in braking index of similar magnitude following this radiatively loud glitch, with a $\sim15\%$ reduction in $n$ at the time of the glitch \citep{2015MNRAS.447.3924A}.
However, only formal phase-connected timing errors are given for this possible change in the braking index, and this method is susceptible to timing noise, \citep[e.g.][]{2010MNRAS.402.1027H, 2011ApJ...730...66L}.
Given this, and the large non-white residuals seen after the fitting, the true significance of this result is currently unknown.

It is interesting  that the only two nominally rotation-powered pulsars which have been observed to have radiatively loud glitches are two of those with the highest dipole-inferred magnetic field.
In both cases the observed braking indices were consistent with being constant through radiatively quiet glitches and decreased following their loud glitches.
This decrease in braking index effectively has the pulsars moving faster towards the magnetar population on the $P$-$\dot{P}$ diagram.
This, together with radiatively loud glitches being a defining characteristic of magnetars \citep[e.g.][]{2014ApJ...784...37D} is suggestive that the large magnetic field in these two seemingly rotation-powered pulsars is responsible for their unusual activity.

There was also a change in $\ddot{\nu}$ in the high-magnetic-field rotation-powered pulsar PSR$\;$J1718$-$3718 following a large glitch \citep{2011ApJ...736L..31M}.
While the implied  $\ddot{\nu}$ both before and after this pulsar's glitch gives nonphysical braking indices, $n\sim-17(5)$ and $n\sim-146(2)$, the measured $\ddot{\nu}$ were consistent over $\sim3000\;$days before the glitch, and for the $\sim700\;$days after it.
Again, while the implied braking indices seem nonphysical, it is interesting that $\ddot{\nu}$ changed with a glitch in yet another high-magnetic-field pulsar.

A possible change in the braking index was seen in the Crab pulsar, where for a $\sim 11$-yr span the measured braking index was $\sim8\%$ lower than the long-term average braking index.
This period of low braking index occurred during a period of higher-than-normal glitch activity, and \cite{2015MNRAS.446..857L} note that this possible change is most likely due to unmodeled glitch parameters.

One possibility to explain a substantial change in a braking index, such as the one we observe in \source{}, would be contamination due to a long-term glitch recovery.
If this is the case, one would expect a bias towards a higher $\ddot{\nu}$, and thus a higher $n$  \citep{2000MNRAS.315..534L}.
This is due to the typical glitch behavior of an exponentially decaying $\nu$, which leads to a decrease in the magnitude of the measured $\dot{\nu}$ as a function of time, and thus to an artificially larger braking index.
This is the opposite of what we observe.

There are several  theoretical models to explain the observation that all measured braking indices are less than the canonical $n=3$ of a magnetic dipole in a vacuum.
As yet, the change in braking index observed in \source{} is unique -- it is larger than ever before seen, and appears to be constant following the magnetar-like event.
Here we will discuss the consequences of a changing braking index in the context of these models.

In particle-wind models \citep[see e.g.][]{1999ApJ...525L.125H, 2013ApJ...768..144T}, one can explain any braking index between  $n=1-3$ by combining spin-down effects from both the standard magnetic dipole radiation ($n=3$) with that of angular momentum loss from an out flowing particle wind ($n=1$).
As shown in \cite{2015MNRAS.446..857L}, one can express the fraction of spin-down power due to a particle wind as:
\begin{equation}
\epsilon=\frac{3-n}{n-1}.
\end{equation}
This would imply that before 2006, 21$\pm$1\% of \source{}'s spin-down was due to a wind, and $68\pm4\%$ after 2006.
This model predicts a relation between the braking index and the luminosity of the particle wind \citep{1999ApJ...525L.125H, 2011ApJ...730...66L}:
\begin{equation}
L_p=(3-n)^2\left(\frac{\dot{\nu}}{\nu}\right) \frac{6I^2c^3}{B^2R^6}
\end{equation}
where $I$ is the moment of inertia, $B$ the magnetic field, and $R$ the pulsar's radius.
Assuming neither the magnetic field nor moment of inertia changed substantially, the luminosity of the pulsar wind nebula might have been expected to increase by a factor of approximately 5.
Such a significant flux change was ruled out by deep {\it Chandra} observations by \cite{2011ApJ...730...66L}, as well as by the consistency of the flux during the {\it Swift} campaign with the pre-outburst flux reported by \cite{2008apj...678.1L43K} to within the telescopes' cross calibration uncertainties, $\sim$10\%.

One can also obtain a braking index different from 3 by relaxing the assumption of a constant magnetic dipole in a vacuum, allowing the dipole to change over time, \cite[see e.g.][]{1969Natur.221..454G, 1985Natur.313..374M, 1988MNRAS.234P..57B}.
This is expressed in a convenient form by \cite{2015MNRAS.446..857L}:
\begin{equation}
n_{obs}=n_{dip}+\frac{\nu}{\dot{\nu}}\left(-\frac{\dot{I}}{I}+2\frac{\dot{\alpha}}{\mathrm{tan}\alpha}+2\frac{\dot{M}}{M}\right).
\end{equation}
To explain a braking index lower than $n_{dip}$ of 3, either the moment of inertia $I$ is decreasing, or either the mass $M$ or the angle of mis-alignment between the spin and magnetic axis $\alpha$ are increasing.
Furthermore, this implies that at the epoch of the magnetar outburst, the fractional rate of change of the magnitude of either $\dot{I}/I$, $\dot{\alpha}/$tan$\alpha$, or $\dot{M}/M$ increased by a factor of 2.3$\pm0.2$.
It does not seem physically plausible to have so large a change in either $\dot{I}/I$ nor $\dot{M}/M$, especially given the lack of change of the flux of the pulsar wind nebula.
Such a change in either $\dot{\alpha}$ or $\alpha$ also seems improbable, given the lack of any detected change in the pulse profile (see \S\ref{subsect:profile}).

One could also change the braking index by altering the geometry of the magnetosphere \citep[see e.g.][]{2002ApJ...574..332T, 2006ApJ...643.1139C}.
In the twisted neutron-star magnetosphere model of \cite{2002ApJ...574..332T}, the braking index of a pulsar is given by $n=2p+1$, where $p$ is radial index.
The observed change in braking index in this model implies that for \source{}, the ``twist'' between the north and south hemispheres increased by $\sim$ one radian at the time of the outburst, which should lead to a corresponding increase in the X-ray luminosity of $\sim50\%$.
This is not seen.
Additionally, in any magnetospheric origin for a change in braking index, one would need to modify the magnetosphere in such a way as to maintain a constant pulse profile over the magnetar-like event, which seems challenging.

\cite{2009ApJ...703.1044B} has a modified version of this model in which instead of a global twist in the magnetosphere, the twist is concentrated into a localized region known as a ``$j$-bundle.''
This $j$-bundle will increase the dipole moment of the neutron star, leading to an increased spin-down rate.
As the $j$-bundle shrinks, the effective dipole moment should decrease with time, leading to a positive contribution to $\ddot{\nu}$ and thus the braking index.
While this model can be used to explain the glitch behavior of \source{} associated with the magnetar-like event \citep{2010ApJ...710.1710L}, it does not immediately explain our observed long-term decrease in braking index.

\section{Conclusions}
The observed braking index of \source{} has significantly changed following its period of magnetar-like behavior.
This long-term change in $n$ is, to within measurement errors, unaccompanied by any corresponding long-lived change in the flux of the source, or any change in its pulse profile.
This is in contrast to most of the models discussed above where a correlated change in the X-ray luminosity is expected for both wind-based models \citep{1999ApJ...525L.125H} and global magnetospheric twist based models \citep{2002ApJ...574..332T}.
As well, models which modify the assumptions of a constant magnetic dipole require far too high  a change in $I$ or $M$  to be physically plausible, or a change in $\alpha$ or $\dot{\alpha}$ which seem unlikely given the stable pulse profile.

The most plausible explanation for a changed braking index appears to be due to some form of change in magnetospheric configuration, but this change is constrained by our observations to be unaccompanied by any large-scale change in flux, spectrum or pulse profile.
One possible way to probe the magnetosphere of pulsars, and therefore test this hypothesis, is by means of X-ray polarimetry.
Measurements of polarization fractions and angles are very sensitive to viewing geometries, as well as twists in the magnetosphere \citep[e.g.][]{2006MNRAS.373.1495V, 2014MNRAS.438.1686T}.

\section{Acknowledgments}
R.F.A. acknowledges support from an  NSERC  Alexander  Graham  Bell  Canada  Graduate  Scholarship and a Walter C. Sumner Memorial Fellowship.
V.M.K. receives support from an NSERC Discovery Grant and Accelerator Supplement, Centre de Recherche en Astrophysique du Quebec, an R. Howard Webster Foundation Fellowship from the Canadian Institute for Advanced Study, the Canada Research Chairs Program and the Lorne Trottier Chair in Astrophysics and Cosmology.
We thank R. Ferdman and E. Madsen for useful discussions. We acknowledge the use of public data from the {\it Swift} data archive.
This research has made use of data obtained through the High Energy Astrophysics Science Archive Research Center Online Service, provided by the NASA/Goddard Space Flight Center.

\bibliography{/homes/borgii/rarchiba/Papers/rfa}{}
\bibliographystyle{apj}

\end{document}